\documentclass[prl,twocolumn,showpacs,amsmath,amssymb,superscriptaddress,a4]{revtex4}
\usepackage{graphicx}
\usepackage{amsmath}
\usepackage{amssymb}
\usepackage{dcolumn}
\usepackage{bm}
\usepackage{pstricks}
\usepackage{psfrag}

\usepackage{color}

\definecolor{ngreen}{rgb}{0.2,0.5,0.2}

\begin{document}

\newcommand{\Chi}{{\rm X}}

\newcommand{\fig}[1]{Fig.~\ref{#1}}
\newcommand{\beq}{\begin{equation}}
\newcommand{\eeq}{\end{equation}}
\newcommand{\bqa}{\begin{eqnarray}}
\newcommand{\eqa}{\end{eqnarray}}
\newcommand{\nn}{\nonumber}
\newcommand{\nl}{\nn \\ &&}
\newcommand{\dg}{^\dagger}
\newcommand{\rt}[1]{\sqrt{#1}\,}
\newcommand{\erf}[1]{Eq.~(\ref{#1})}
\newcommand{\erfs}[1]{Eqs.~(\ref{#1})}
\newcommand{\erfand}[2]{Eqs.~(\ref{#1}) and (\ref{#2})}
\newcommand{\Erf}[1]{Equation~(\ref{#1})}
\newcommand{\smallfrac}[2]{\mbox{$\frac{#1}{#2}$}}
\newcommand{\bra}[1]{\left\langle{#1}\right|}
\newcommand{\ket}[1]{\left|{#1}\right\rangle}
\newcommand{\ip}[2]{\left\langle{#1}\right|\left.{#2}\right\rangle}
\newcommand{\sch}{Schr\"odinger}
\newcommand{\hei}{Heisenberg}
\newcommand{\half}{\smallfrac{1}{2}}
\newcommand{\bl}{{\bigl(}}
\newcommand{\br}{{\bigr)}}
\newcommand{\ito}{It\^o }
\newcommand{\sq}[1]{\left[ {#1} \right]}
\newcommand{\cu}[1]{\left\{ {#1} \right\}}
\newcommand{\ro}[1]{\left( {#1} \right)}
\newcommand{\an}[1]{\left\langle{#1}\right\rangle}
\newcommand{\tr}[1]{{\rm Tr}\sq{ {#1} }}
\newcommand{\st}[1]{\left| {#1} \right|}

\title{Adaptive Optical Phase Estimation Using Time-Symmetric Quantum Smoothing}
\author{T. A. Wheatley}
\affiliation{Centre for Quantum Computer Technology, Australian Research Council}
\affiliation{School of Engineering and Information Technology, University College, The University of New South Wales, Canberra 2600, ACT, Australia}
\affiliation{Department of Applied Physics and Quantum Phase Electronics Center, School of Engineering, The University of Tokyo, 7-3-1 Hongo, Bunkyo-ku, Tokyo 113-8656, Japan}
\author{D. W. Berry}
\affiliation{Institute for Quantum Computing, University of Waterloo, Waterloo, ON, Canada }
\author{H. Yonezawa}
\author{D. Nakane}
\author{H. Arao}
\affiliation{Department of Applied Physics and Quantum Phase Electronics Center, School of Engineering, The University of Tokyo, 7-3-1 Hongo, Bunkyo-ku, Tokyo 113-8656, Japan}
\author{ D. T. Pope}
\affiliation{Perimeter Institute for Theoretical Physics, 31 Caroline Street N, Waterloo, ON N2L 2Y5, Canada}
\author{T. C. Ralph}
\email{ralph@physics.uq.edu.au}
\affiliation{Centre for Quantum Computer Technology, Australian Research Council}
\affiliation{Department of Physics, University of Queensland, Brisbane 4072, QLD, Australia}
\author{H. M. Wiseman}
\email{h.wiseman@griffith.edu.au}
\affiliation{Centre for Quantum Computer Technology, Australian Research Council}
\affiliation{Centre for Quantum Dynamics, Griffith University, Brisbane 4111, QLD, Australia }
\author{A. Furusawa}
\email{akiraf@ap.t.u-tokyo.ac.jp}
\affiliation{Department of Applied Physics and Quantum Phase Electronics Center, School of Engineering, The University of Tokyo, 7-3-1 Hongo, Bunkyo-ku, Tokyo 113-8656, Japan}
\author{E. H. Huntington}
\email{e.huntington@adfa.edu.au}
\affiliation{Centre for Quantum Computer Technology, Australian Research Council}
\affiliation{School of Engineering and Information Technology, University College, The University of New South Wales, Canberra 2600, ACT, Australia}
\begin{abstract}
Quantum parameter estimation has many applications, from gravitational wave detection to quantum key distribution.  We present the first experimental demonstration of  the time-symmetric technique of quantum smoothing.  We consider both
adaptive and non-adaptive quantum smoothing, and show that both are better than their well-known time-asymmetric counterparts (quantum filtering). For the problem of estimating a stochastically varying phase shift on a coherent beam, our theory predicts that adaptive quantum smoothing (the best scheme) gives an estimate with a mean-square error up to $2\sqrt{2}$ times smaller than that from non-adaptive quantum filtering (the standard quantum limit). The experimentally measured improvement is $2.24 \pm 0.14$.
\end{abstract}
\pacs{42.50.Dv, 42.50.Xa, 03.65.Ta, 03.67.-a, 06.90.+v}
\maketitle

\newpage
 Quantum parameter estimation (QPE) is the problem of estimating an unknown classical parameter (or process) which plays a role in the preparation (or dynamics) of a quantum system \cite{Giovannetti_Q_meas,WisMil09}, and is central to many fields including gravitational wave interferometry \cite{gw},  quantum computing \cite{hofheinz_QStates}, and quantum key distribution  \cite{DPS_QKD}. The fundamental limit to the precision of the estimate in QPE is set by quantum mechanics \cite{Giovannetti_Q_meas,WisMil09}. Thus one of the key issues in QPE is the development of practical methodologies which allow measurements to approach or exceed the standard quantum limit (SQL) for a given measurement coupling  \cite{Geremia03,Silberhorn_det_Q_light,vanDam2007,LupascuQND,Branczyk2007,anderson_exp_st_est,cook_state_discr}.  Because of its wide-ranging technological relevance,
the prime example of QPE is estimating an optical phase shift \cite{SumPeg90,SandMil95,wiseman_PRL_75,BerWis00,mabuchi,Berry,Pryde,Tsang_PLL}.

Apart from some theoretical papers \cite{Berry,Tsang_PLL}, work in this area of QPE has concentrated
upon the problem of estimating a {\em fixed}, but unknown phase shift, which can be thought of as preparing the quantum state with an average phase equal to this parameter. It was shown theoretically \cite{wiseman_PRL_75} that for this problem {\em adaptive} homodyne measurements coupled with an optimal estimation filter can yield an estimate with mean-square error smaller than the standard quantum limit (as set by perfect heterodyne detection).
This was demonstrated experimentally in Ref. \cite{mabuchi} using very weak coherent states (for which
the factor of improvement is at most 2). More recent theory and experiment have shown that interferometric measurements with photon counting can also be improved using adaptive techniques \cite{BerWis00,Pryde}.

A far richer, and in many cases more experimentally relevant, problem of quantum phase estimation arises when the phase {\em evolves} dynamically under the influence of an unknown classical stochastic process \cite{Berry,Tsang_PLL}. The general
problem of estimating a classical process dynamically coupled to a quantum system under continuous measurement has recently been considered by Tsang \cite{Tsang_smooth}, who introduced
three main categories of quantum estimation: prediction or filtering, smoothing, and retrodiction.
 Of those, prediction or filtering is a causal estimation technique that can be used in real-time applications \footnote{Tsang \cite{Tsang_smooth} makes the distinction that prediction uses data taken prior to the time of the estimate, whereas filtering uses data taken up to the time of the estimate.}.
 Smoothing and retrodiction are acausal and so cannot be used in real time, but they can be used for off-line data processing or with a delay corresponding to the estimation time.   Smoothing, in which the signal is inferred at a point in time based on data taken both before and after that time, is the only time-symmetric estimation technique. As a consequence, it can be more precise than the time asymmetric techniques of filtering or  retrodiction \cite{Tsang_PLL,Tsang_smooth}.   Such a result is very significant for quantum sensing applications where it is more important to have precise rather than real-time estimates.

Here we present the first experimental demonstration of QPE using quantum smoothing.   Specifically, we  consider estimation of the phase of a continuous optical field, generalizing the theory of Ref.~\cite{Berry} to a more general classical phase noise process (rather than pure diffusion), and to smoothing (rather than filtering).  According to our theory, adaptive measurements and smoothing both offer improvements over the alternative (non-adaptive and filtering respectively). Moreover, using both together offers the maximum improvement, with a mean-square phase error smaller than the standard (non-adaptive, filtered) quantum limit by a factor of up to $2\sqrt{2}$ in theory for pure phase diffusion. We verify these predicted improvements experimentally, for the first time in every case, and find a maximum improvement by a factor of $2.24\pm 0.14$ over the SQL.

\begin{figure}[!htbp]
   \includegraphics[width=\columnwidth]{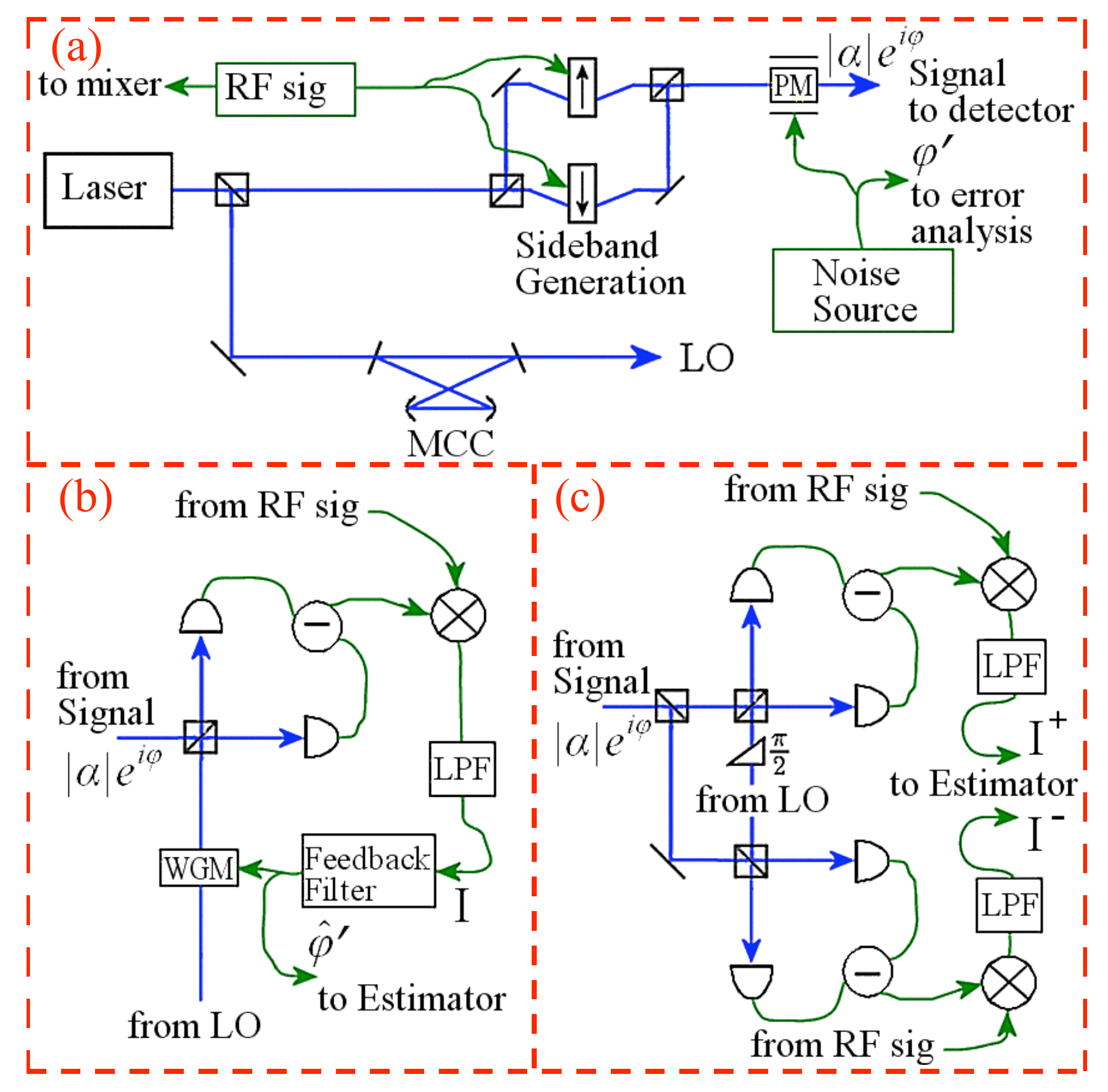}
 \caption{\label{comb_setup} Schematic diagrams showing: (a) source and local oscillator generation; (b) adaptive phase estimation; (c) dual homodyne phase estimation.  Although illustrated as a single device, both AOMs as drawn are actually a pair of AOMs which shift by $110$ MHz and $105$ MHz in opposite directions to achieve a $5$ MHz frequency shift.  LO = local oscillator; RF sig = radio-frequency signal; PM = phase modulator; WGM = waveguide modulator; LPF = low-pass filter; MCC = mode-cleaning cavity; AOM = acousto-optic modulator.}
   \end{figure}

Fig.~\ref{comb_setup} illustrates the system under consideration.   The goal in this quantum sensing problem is to form the optimal estimate $\Theta(t)$ of the system phase $\varphi(t)$ of a weak coherent state in the presence of noise in the measurement and classical noise in the system  phase.  The precision of the estimate is quantified by the mean-square error between the estimate and the actual phase such that $\sigma^2 \equiv \langle [\Theta(t) - \varphi(t)]^2 \rangle$.

 Unlike previous adaptive phase estimation experiments \cite{mabuchi,Pryde},  we compare the phase estimate to the actual system phase in order to directly measure the error in estimation.   This is achieved by deliberately imposing classical phase noise via an electro-optic phase modulator (PM), as indicated in Fig.~\ref{comb_setup}(a).  A titanium:sapphire laser operating at 860nm is used to drive the experiment.  The arrangement of acousto-optic modulators (AOMs) shown in Fig.~\ref{comb_setup}(a) is used to generate a pair of phase modulation sidebands at 5 MHz such that it is equivalent to a weak coherent state with a photon flux ${\cal N} = |\alpha|^2$ of order $10^6$ photons per second.

The phase noise is applied using a PM driven by an Ornstein-Uhlenbeck (OU) \cite{Gar04} noise source. The phase variation is
\begin{equation}
\label{ou}
d\varphi(t) = -\lambda \varphi(t) dt + \sqrt{\kappa} dV(t),
\end{equation}
where $dV$ is a Wiener increment and $\kappa$ is the inverse coherence time. For this experiment $\kappa$ is of order $10^4$~rad/s, so there are about 100 photons per coherence time. We record the imposed voltage at the monitor port of the high voltage amplifier ($\varphi^\prime$ in Fig.\ref{comb_setup}).  The phase deviation arising from that voltage is calibrated via the half-wave voltage of the PM and used as our measure of the true system phase $\varphi$.

 An arbitrary  quadrature of the field of interest can be measured with a balanced homodyne detector, in which the field of interest is interfered with a 1.5 mW local oscillator on a beamsplitter.  Both outputs of the beamsplitter are detected and the resulting measurements are subtracted to form the homodyne photocurrent $I(t)$ \cite{BachorRalph}.   The detection efficiency (including homodyne fringe visibility  of 97\%, detector quantum efficiency of 98\% , and optical transmission of  97\% ) was in excess of 89\% in all measurements and the electronic noise floor was  11~dB below the shot noise of the measurements.  In all cases, the homodyne detector is DC-locked to ensure that the deliberately imposed OU noise dominates  the uncertainty in the phase.

The adaptive phase estimation system is illustrated in Fig.~\ref{comb_setup}(b).  The output  of the homodyne detector is electronically demodulated to give $I(t)$ which is then fed into the feedback filter. This yields a voltage $\hat\varphi'$, which is stored for later data analysis and also fed back so as to imprint a phase $\hat\varphi \propto \hat\varphi'$ on the optical LO using a WGM. That is, the {\em intermediate phase estimate} $\hat\varphi$ is the phase of the measured quadrature.

Because $\varphi \approx \hat\varphi$, we can use a  linearized approximation for the homodyne photocurrent:
\begin{equation}
I(t)dt = 2|\alpha|[\varphi(t)-\hat\varphi(t)]dt + dW(t),
\end{equation}
where $dW(t)$ is Wiener noise arising from the quantum vacuum fluctuations.
We define the {\em instantaneous estimate} $\theta(t)$ to the best estimate of $\varphi(t)$ which can be made using only the data taken in the time interval $[t,t+dt)$:
\begin{eqnarray}
\theta(t)&:=&\hat\varphi(t)+I(t)/2|\alpha| \label{thetadapt1} \\
&=&\varphi(t)+\frac{dW(t)}{dt}\frac{1}{2\sqrt{\cal N}} \label{thetadapt2}.\label{estimator}
\end{eqnarray}
To obtain an intermediate estimate $\hat\varphi$ with a finite amount of noise it is necessary to
time-average the instantaneous estimate. This can be achieved by using a simple integrator on
$I(t)$~\cite{Berry}, but for practical reasons we use a linear low-pass filter:
\beq \label{hatvarphi}
\hat\varphi(t) = \int_{-\infty}^{t} \beta e^{\omega_0 (s-t)} I(s)/2\sqrt{\cal N} ds.
\eeq
We work in the limit where the cut-off frequency $\omega_0 \approx 10^2$s$^{-1}$ is much less than
the feedback gain $\beta > 10^5$s$^{-1}$.  In this limit,
it is easy to verify from the above equations that the intermediate estimate reduces to the filter used in Ref.~\cite{Berry}:
\beq
\hat\varphi(t) =  \int_{-\infty}^{t}  \beta e^{\beta(s-t)} \theta(s) ds.
\label{hatvarphi2}
\eeq

The intermediate estimate $\hat\varphi(t)$ of $\varphi(t)$
is always a filtered estimate, because
it is used in a causal feedback loop. The theory of Ref.\ \cite{Berry} also used filtering to obtain the
{\em final phase estimate} for $\varphi(t)$. That is, the final estimate for time $t$
was based only on the data
obtained up to time $t$. However, it is possible to obtain a better phase estimate
by smoothing: using the data after time $t$ also. Here we can
assume that data for an infinite period of time before and after $t$ can be used,
because the experimental data run time of $10^{-2}$s is long compared to the averaging time,
which is $\lesssim 10^{-4}$s.

Let us denote by $\Theta_-(t)$ and $\Theta_+(t)$ the phase
estimates obtained from data obtained before and after time $t$, respectively.
Following  Ref.\ \cite{Berry}, we consider estimates that are weighted averages of the {\em instantaneous estimates}:
\begin{equation}
\Theta_{\pm}(t) = \pm \chi_{\pm} \int_{t}^{\pm \infty} \theta(s)
e^{\mp  \chi_{\pm}(s-t)} ds.
\label{est_F_B}
\end{equation}
The deviation of these from the actual phase is then \begin{align}
\label{eq:min}
\Theta_{\pm}(t) - \varphi(t)
&=  \pm \chi_{\pm} \int_{t}^{\pm \infty}
e^{\mp  \chi_{\pm}(s-t)} [\varphi(s) - \varphi(t)]ds \nonumber \\ & \quad  \pm \frac{\chi_{\pm} } {2 \sqrt{N}}  \int_{t}^{\pm \infty} e^{\mp  \chi_{\pm}(s-t)} dW(s).
\end{align}
The forwards estimate, $\Theta_-(t)$ corresponds to the causal (or filtered) estimate. A weighted average of the forwards and backwards estimates can be used to construct the time-symmetric (or smoothed) estimate $\Theta (t) = w_-\Theta_-(t)+w_+\Theta_+(t)$, the variance of which is:
\begin{eqnarray}
\label{smoothedEstimate}
\sigma^2&=&w_-^2\langle [\Theta_-(t) - \varphi(t)]^2 \rangle
 +w_+^2\langle [\Theta_+(t) - \varphi(t)]^2 \rangle \nonumber \\
&& +2w_-w_+\langle [\Theta_-(t) - \varphi(t)][\Theta_+(t) - \varphi(t)] \rangle.
\end{eqnarray}

This can be evaluated using the definition of the OU process for the
system phase \eqref{ou}.
From the time symmetry of this process,
the mean-square errors in $\Theta_+ (t)$ and $\Theta_- (t)$ are the same:
\begin{equation}
\label{filteredEstimate}
\sigma_{\pm}^2  = \frac{\kappa}{2(\chi_{\pm} +\lambda)}
+\frac{\chi_{\pm}}{8{\cal N}},
\end{equation}
while the correlation between the forwards and backwards estimates is
\begin{equation}
\langle [\Theta_-(t) - \varphi(t)][\Theta_+(t) - \varphi(t)] \rangle =
\frac{\kappa \lambda}{2(\chi_-+\lambda)(\chi_++\lambda)}.
\end{equation}
By symmetry, \erf{smoothedEstimate} expression is minimised for $\chi_-=\chi_+=\chi$ and $w_-=w_+=1/2$, which gives
\begin{align}
\label{eq:ans}
\sigma^2 = \frac{\kappa (\chi+2\lambda)}
{4(\chi+\lambda)^2}+\frac{\chi}{16{\cal N}}.
\end{align}
Equations (\ref{filteredEstimate}) and (\ref{eq:ans}) are simplified greatly in the limit $\xi \equiv \lambda / (2\sqrt{\kappa {\cal N}}) \ll 1$,
which is a good guide for our experiment where $\xi \approx 0.2$.
In this limit, the optimal value of $\chi$ in \erf{filteredEstimate} is $2\sqrt{\kappa {\cal N}}$, giving the minimum variance $\sigma_-^2 = \sqrt{\kappa/{\cal N}}/2$. The relative  corrections are  $O(\xi^2)$. Changing to smoothing reduces the variance by a factor of 2 to $\sigma^2 = \sqrt{\kappa/{\cal N}}/4$.

We compare the results of the adaptive technique to the standard technique for phase estimation,
dual homodyne detection, illustrated in Fig.~\ref{comb_setup}(c). It incurs the same noise penalty as heterodyne detection.  The dual homodyne data [$I^+$ and $I^-$ in Fig.~\ref{comb_setup}(c)] can be used to form an instantaneous estimate, comparable to \erf{thetadapt1}, via
\beq
\theta_{\rm s} (t)= {\rm arg}[I_+(t) + iI_-(t)].
\eeq
This can be shown to give an estimate which is effectively the same as \erf{thetadapt2},
but with an additional noise penalty incurred from measuring both quadratures: ${\cal N}$
must be replaced by ${\cal N}_{\rm s} = {\cal N}/2$ \cite{Berry}. The mean-square errors for dual homodyne measurements are simply obtained by making this substitution in Eqs.~(\ref{filteredEstimate}) and (\ref{eq:ans}). Comparing these to the adaptive results, the latter thus give a  reduction in the variance by a factor of $\sqrt{2}$ for   both filtering and smoothing. The SQL is set by nonadaptive filtering, and equals   $\sigma_{s-}^2 = \sqrt{\kappa/{\cal N}_{\rm s}}/2$.

 \begin{figure}[!htbp]
   \includegraphics[width=\columnwidth]{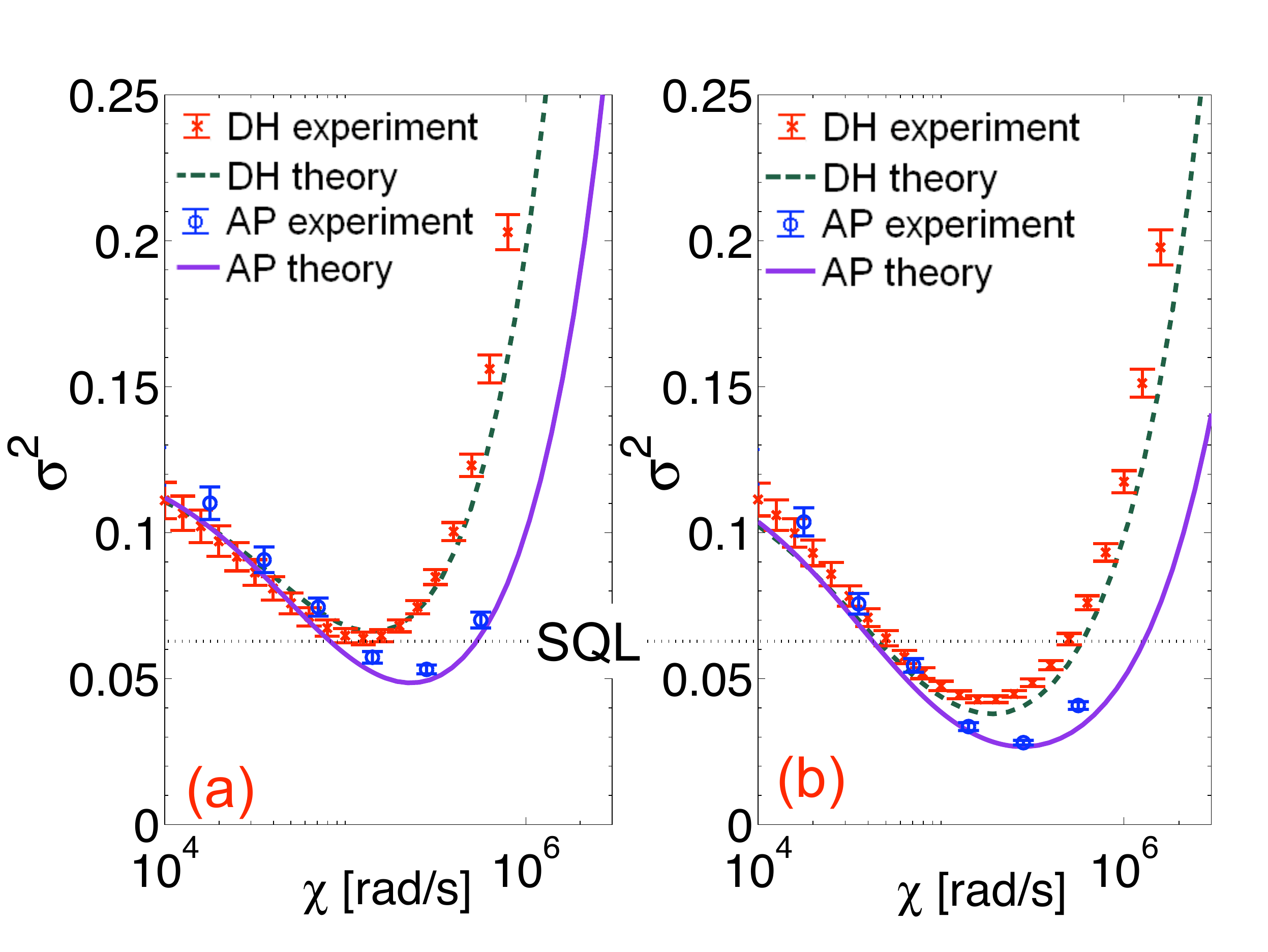}
 \caption{\label{results} The experimental and theoretical variance $\sigma^2$ of the four phase estimation techniques: filtered dual homodyne (DH) and adaptive phase (AP) in part (a); and smoothed DH and AP in part (b).  Parameters are: $\kappa_{DH}=1.6218\times10^4$~rad/s, $\lambda_{DH}=6.4593\times10^4$ rad/s, ${\cal N}_{DH}=1.3235\times 10^6$ s$^{-1}$, $\kappa_{AP}=1.5868\times10^4$ rad/s, $\lambda_{AP}=6.1451\times10^4$ rad/s, ${\cal N}_{AP}=1.3499\times 10^6$ s$^{-1}$. }
   \end{figure}

The measured and predicted mean-square errors for the four different estimation techniques are shown in Fig.~\ref{results}.  The values of $\kappa$ and $\lambda$ are determined from the calibrated measurements of the system phase $\varphi(t)$.  The photon number, $ {\cal N} =|\alpha^2|$,  is from the measured amplitude of the coherent state relative to the quantum noise limit, while $\chi$ is varied in the experiment as indicated in Fig.~\ref{results}.  Each data run is 10 ms long, and error bars are the standard deviation of multiple data sets.  We performed the adaptive estimates  $\Theta_\pm (t)$ not by averaging $\theta(t)$ as in
\erf{est_F_B}, but rather by averaging $\hat\varphi(t)$. This gives more stable results, and
is justified since, in the limit $\xi \ll 1$, the optimal value of $\beta$ in \erf{hatvarphi} is $\sqrt{8\chi{\cal N}}$ \cite{Berry}. In the regime of the experiment, this $\beta$ is much greater than $\chi$, so the extra averaging in \erf{hatvarphi2} is negligible.

Figure~\ref{results} shows good agreement between theory and experiment for all the estimation techniques. It demonstrates that phase estimation by quantum smoothing is significantly better than that from quantum filtering. As predicted, the improvement is nearly a factor of two at the optimum value of $\chi$ for both the adaptive and dual homodyne measurements. Figure~\ref{results} also shows the first experimental verification of the quantum theory of continuous adaptive phase estimation \cite{Berry}.  As predicted, adaptive phase estimation outperforms dual homodyne measurement by a factor of approximately $\sqrt{2}$. The theory curves here take into account the known imperfections (detector dark noise, homodyne efficiency and optical transmission losses). However, the horizontal line indicating the SQL $\sigma^2_{s-}$ is defined (as above) in terms of the actual photon flux, and corresponds to what would be achievable by {\em ideal} dual-homodyne filtering.  Note that  adaptive measurements perform better than the SQL for both types of estimator.

   \begin{figure}[!htbp]
   \includegraphics[width=\columnwidth]{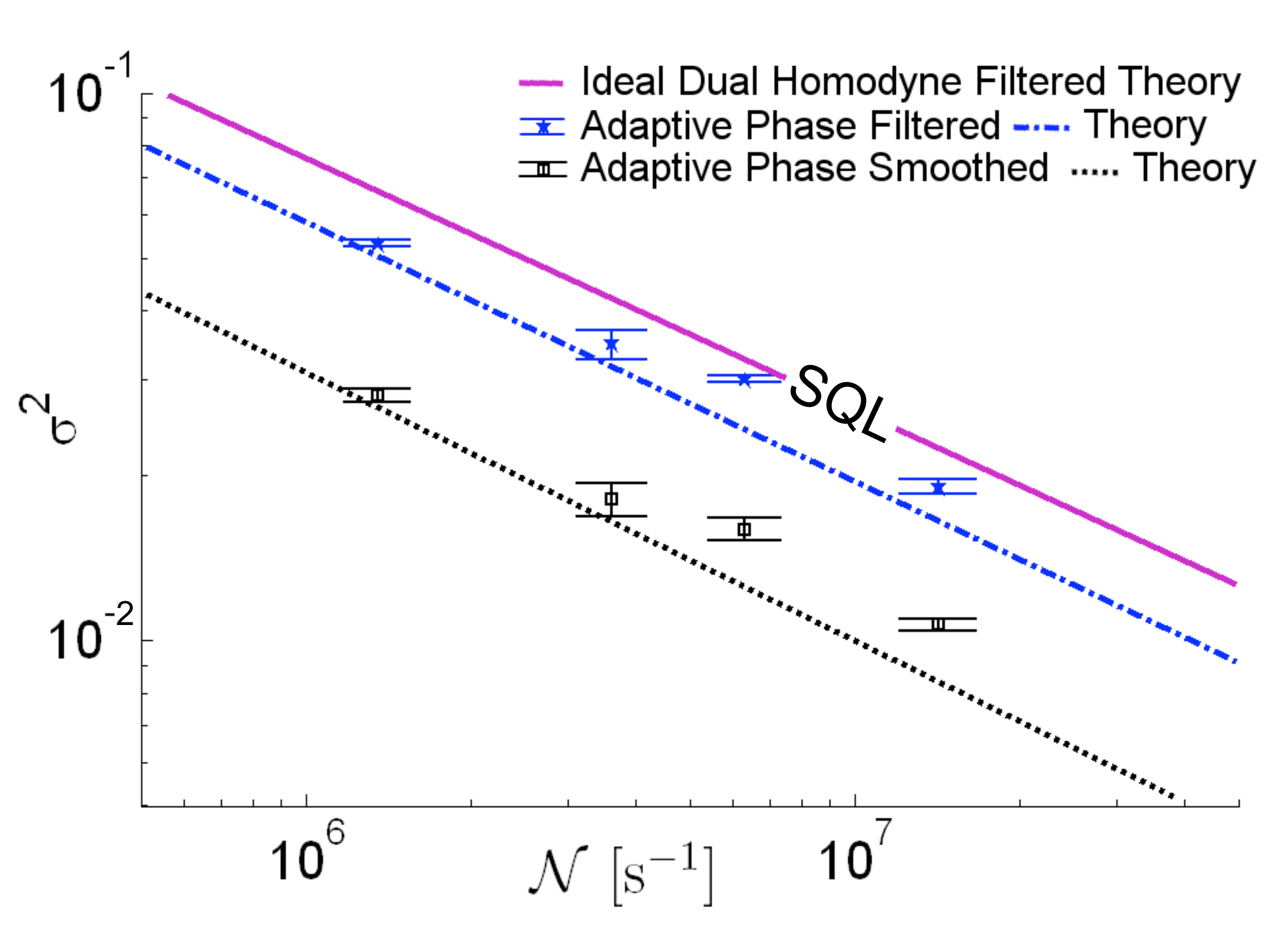}
 \caption{\label{resultsN} The variance $\sigma^2$ of the adaptive phase estimation for quantum filtering and smoothing as a function of photon number ${\cal N}$, compared to the relevant theoretical predictions, and the
 theoretical predictions for nonadaptive measurements.}
  \end{figure}

Figure~\ref{resultsN} shows the optimal mean-square errors in the filtered and smoothed adaptive estimates for four different values of measured photon number, spanning an order of magnitude between  ${\cal N} \approx 10^6$~s$^{-1}$ and ${\cal N} \approx 10^7$~s$^{-1}$.  The results confirm that quantum smoothing outperforms quantum filtering over a wide range of photon numbers, as predicted by theory.  As in Fig.~\ref{results}, the SQL is set by ideal dual homodyne filtering, while the other theory lines are for the actual (non-ideal) experiment.  As predicted, adaptive measurements out-perform the SQL.

In summary, we have demonstrated experimentally and theoretically that estimation of the phase of an optical field in the presence of classical noise using quantum smoothing is superior to the equivalent quantum filtered approach.   We have also demonstrated experimentally for the first time that continuous adaptive measurements perform better than the standard quantum limit for both types of estimator.  Combining quantum smoothing with adaptive measurements gives the maximum improvement over the standard (perfect non-adaptive, filtering) quantum limit. The experimental improvement of a factor of $2.24 \pm 0.14$ in the mean-square error compares well with the theoretical maximum (using phase diffused coherent states) of $2\sqrt{2}$. These insights and techniques will be applicable to the even more interesting case of estimation using non-classical states, where the improvement can be arbitrarily large.

This work was supported financially by SCF, GIA, G-COE, PFM of MEXT, REFOST,
SCOPE of MIC, the ARC, Industry Canada and the Province of Ontario.

\bibliographystyle{prsty}                      

\begin{thebibliography}{99}
\bibitem{Giovannetti_Q_meas}
V.~Giovannetti, S.~Lloyd, and L.~Maccone,
Science {\bf 306}, 1330 (2004).

\bibitem{WisMil09}
H.~M. Wiseman and G.~J. Milburn, \textit{Quantum Measurement and Control},  (Cambridge University Press, Cambridge, 2010)

\bibitem{hofheinz_QStates}
M.~Hofheinz
 \ \textit{et al.},
\newblock Nature {\bf 459}, 546, (2009).

\bibitem{DPS_QKD}
K.~Inoue, E.~Waks, and Y.~Yamamoto.
\prl {\bf 89}, 037902 (2002).

\bibitem{gw}
K. Goda \textit{et al.}, Nature Physics {\bf 4}, 472 (2008).

\bibitem{Geremia03}
J.~M. Geremia, J.~K. Stockton, A.~C. Doherty, and H. Mabuchi, \prl \textbf{91}, 250801 (2003).

\bibitem{Silberhorn_det_Q_light}
C.~Silberhorn, Contemporary Physics {\bf 48}, 143 (2007).

\bibitem{vanDam2007} W. van Dam
\ \textit{et al.},
\prl \textbf{98}, 090501 (2007).

\bibitem{LupascuQND}
A.~Lupascu
  \ \textit{et al.}, Nature Physics {\bf 3}, 119 ( 2007).

\bibitem{Branczyk2007}
A.~M. Bra\'nczyk%
\ \textit{et al.}%
, \pra \textbf{75}, 012329 (2007).

\bibitem{anderson_exp_st_est}
U.~L. Andersen, M.~Sabuncu, R.~Filip, and G.~Leuchs, \prl {\bf 96}, 020409 (2006).

\bibitem{cook_state_discr}
R.~L. Cook, P.~J. Martin, and J.~M. Geremia, Nature {\bf 446}, 774 (2007).

\bibitem{SumPeg90} G. S. Summy and D. T. Pegg, Opt. Comm. \textbf{77}, 75 (1990).

\bibitem{SandMil95} B. C. Sanders and G. J. Milburn, Phys. Rev. Lett. {\bf 75}, 2944 (1995).

\bibitem{wiseman_PRL_75}
H.~M. Wiseman, \prl {\bf 75}, 4587 (1995); H. M. Wiseman and R. B. Killip,
\pra {\bf 56}, 944 (1997); and H. M. Wiseman and R. B. Killip,
\pra {\bf 57}, 2169 (1998).




\bibitem{mabuchi}
M. A. Armen
\ \textit{et al.}, \prl {\bf 89}, 133602 (2002).

\bibitem{BerWis00}
D. W. Berry and H. M. Wiseman, \prl {\bf 85}, 5098 (2000).

\bibitem{Pryde}
B.~L. Higgins
\ \textit{et al.},
\newblock Nature {\bf 450}, 393 (2007); and B.~L. Higgins%
\ \textit{et al.}%
, New J. Phys. \textbf{11}, 073023 (2009).


\bibitem{Berry}
D.~W. Berry and H.~M. Wiseman, \pra {\bf 65}, 043803 (2002).

\bibitem{Tsang_PLL}
M.~Tsang, J.~H. Shapiro, and S.~Lloyd, \pra {\bf 79}, 053843 (2009).

\bibitem{Tsang_smooth}
M.~Tsang, \prl {\bf 102}, 250403 (2009).

\bibitem{Gar04}
C. W. Gardiner,
{\em Handbook of Stochastic Methods}
(Spring\-er, Berlin, 2004).

\bibitem{BachorRalph}
H-A. Bachor and T.~C. Ralph,
\newblock {\em A Guide to Experiments in Quantum Optics}.
\newblock Wiley-VCH, Weinheim, 2nd edition, 2004.

\end{thebibliography}

\end{document}